\documentclass[aps,pra,reprint,superscriptaddress,longbibliography,nofootinbib]{revtex4-2}

\usepackage[T1]{fontenc}
\usepackage{lmodern}
\usepackage{amsmath,amssymb,amsfonts,bm,mathtools}
\usepackage{graphicx}
\graphicspath{{./}{./ehe_run_v2/ehe_theory_figures/}{./ehe_revision_figures/}}
\usepackage{xcolor}
\usepackage{physics}
\usepackage{braket}
\usepackage{float}
\usepackage{textcomp}
\usepackage{color}
\usepackage[colorlinks=true,linkcolor=red]{hyperref}
\usepackage[sort&compress]{natbib}
\usepackage[normalem]{ulem}
\usepackage[shortlabels]{enumitem}
\usepackage{appendix}
\usepackage{soul}
\usepackage{amsmath}
\allowdisplaybreaks

\usepackage{siunitx}

\input{epsf}

\begin{document}

\title{Internal cyclotron probe of the vertical polarizability of a surface-state electron on helium}

\author{Kirill Shulga}
\email{kirill\_shulga@protonmail.ch}
\affiliation{International Center for Elementary Particle Physics, The University of Tokyo, 7-3-1 Hongo, Bunkyo-ku, Tokyo 113-0033, Japan}

\begin{abstract}
We study a surface-state electron on liquid helium in a tilted magnetic field. The tilt couples the cyclotron ladder to the vertical Rydberg-like subbands of the same electron and thereby realizes a self-coupled Landau-subband Hamiltonian. In the far-detuned regime, the cyclotron line acquires a state-dependent shift described by standard second-order perturbation theory.
The resulting expression identifies the cyclotron resonance as a direct probe of the dynamical polarizability of the vertical ladder. The theory is intrinsically multilevel. 
Exact diagonalization confirms a broad microwave window in which sizable state-dependent cyclotron shifts are expected while the system remains dispersive.
\end{abstract}

\maketitle

Electrons on liquid helium provide one of the cleanest realizations of a charged quantum particle in a tunable mesoscopic environment \cite{monarkha2004book,platzman1999quantum,lyon2006spin,dykman2003qubits}. The motion normal to the surface forms a Rydberg-like ladder of strongly anharmonic bound states, while a perpendicular magnetic field quantizes the in-plane motion into Landau levels. These two sectors are both experimentally accessible, but they remain physically distinct: one is set by the image-bound surface potential together with the holding field, the other by cyclotron quantization in the plane.

This platform is already connected to several lines of quantum control and readout research. Early work emphasized qubit proposals based on the vertical and spin degrees of freedom \cite{platzman1999quantum,lyon2006spin,dykman2003qubits}, while later studies proposed cavity-based manipulation and readout of trapped electrons \cite{schuster2010proposal}. Resonator-based detection of lateral motion and charge configurations has since been demonstrated for ensembles and for single trapped electrons \cite{yang2016ensemble,abdurakhimov2016strong,chen2018strong,koolstra2019single,castoria2025sensing}, and direct electrical detection of vertical Rydberg transitions has also advanced substantially \cite{kawakami2019imagecharge,belianchikov2025imagecharge,jennings2025quantumcap}. The question addressed here is what changes when the probe is not an external cavity or electrode, but another internal degree of freedom of the same electron.

A tilted magnetic field couples the vertical and cyclotron sectors. On helium, this coupling has already been established in the resonant and near-resonant regime, where it produces renormalization of the vertical transition and, at stronger coupling, dressed-state splitting \cite{yunusova2019coupling,zadorozhko2021motional}. The present work addresses the complementary far-detuned regime, in which the cyclotron line itself becomes the observable of interest. Closely related subband--Landau Hamiltonians have long been studied in finite-width semiconductor systems, where they underlie combined intersubband--cyclotron resonances, tilted-field hybridization, and collective magneto-optical response \cite{ando1979intersubband,merlin1987parabolic,borroff1987raman,observation1988tilted,kumada2008modulation,oh1994threedots}. Electrons on helium differ from those systems in an important respect: the vertical motion is strongly anharmonic and nearly hydrogenic, the single-electron limit is realistic, and the relevant frequencies can be tuned deep into a far-detuned regime.

In that regime the cyclotron line probes a response function of the same electron whose vertical state is being interrogated. More specifically, the tilted field makes the cyclotron degree of freedom an in situ spectroscopic probe of the dynamical polarizability of the vertical Rydberg ladder.

A two-level Rabi model does not provide a complete microscopic description of this system. The interaction is proportional to the physical coordinate operator $z$, which inherits the strong asymmetry of the surface-bound potential. Diagonal and off-diagonal matrix elements of $z$ therefore coexist, and higher vertical subbands are not removed by simple selection rules. The appropriate low-energy theory is generically multilevel.

In this work, we develop the theory directly from the microscopic tilted-field Hamiltonian of a surface-state electron on helium. Using standard second-order perturbation theory, we obtain the state-dependent cyclotron shift, identify it with the dynamical polarizability of the vertical ladder, and show that a generalized Rabi description emerges only after a two-subband projection. Exact diagonalization confirms that the accessible microwave window is broad, that the dispersive regime extends well beyond the strict two-subband limit, and that the cyclotron response gives direct access to the multilevel vertical structure.

%\clearpage

\paragraph*{Microscopic model.}
We consider one electron above the surface of liquid helium in a magnetic field $\mathbf{B}=(0,B_y,B_z)$, where $B_z$ is perpendicular and $B_y$ is parallel to the surface. A holding electric field $E_\perp$ is applied along $z$. The microscopic Hamiltonian is
\begin{equation}
\begin{aligned}
H={}&\frac{p_z^2}{2m_e}+V_{\rm He}(z)+eE_\perp z+\frac{m_e\omega_y^2 z^2}{2} \\
&+\hbar\omega_c\left(a^\dagger a+\frac{1}{2}\right)+\hbar\gamma_B(a+a^\dagger)z,
\end{aligned}
\label{eq:Hmicro}
\end{equation}
with
\begin{equation}
V_{\rm He}(z)=V_0\Theta(-z)-\frac{\Lambda_{\rm im}}{z}\Theta(z),
\label{eq:VHe}
\end{equation}
\begin{equation}
\Lambda_{\rm im}=\frac{\varepsilon-1}{\varepsilon+1}\frac{e^2}{16\pi\varepsilon_0},
\label{eq:LambdaIm}
\end{equation}
and
\begin{equation}
\begin{aligned}
\omega_c&=\frac{eB_z}{m_e}, &
\omega_y&=\frac{eB_y}{m_e}, \\
 l_B&=\sqrt{\frac{\hbar}{eB_z}}, &
\gamma_B&=\frac{\omega_y}{\sqrt{2}\,l_B}.
\end{aligned}
\label{eq:scales}
\end{equation}
Here $p_z$ is the electron momentum normal to the surface, $m_e$ is the electron mass, $V_0$ is the repulsive barrier at the helium surface, $\Theta$ is the Heaviside step function, $\varepsilon$ is the dielectric constant of liquid helium, and $a$ and $a^\dagger$ are the Landau ladder operators. The fourth term in Eq.~(\ref{eq:Hmicro}) is the diamagnetic modification of the vertical confinement produced by the in-plane field, while the last term is the bilinear coupling between the cyclotron coordinate and the vertical motion \cite{yunusova2019coupling,zadorozhko2021motional}.

The vertical part of Eq.~(\ref{eq:Hmicro}),
\begin{equation}
H_z=\frac{p_z^2}{2m_e}+V_{\rm He}(z)+eE_\perp z+\frac{m_e\omega_y^2 z^2}{2},
\qquad
H_z|n\rangle=\epsilon_n|n\rangle,
\label{eq:Hz}
\end{equation}
defines a ladder of bound subbands. Introducing the matrix elements
\begin{equation}
z_{nm}=\langle n|z|m\rangle,
\label{eq:znm}
\end{equation}
the full Hamiltonian becomes
\begin{equation}
\frac{H}{\hbar}=\sum_n\omega_n|n\rangle\langle n|+\omega_c a^\dagger a
+\sum_{n,m}g_{nm}|n\rangle\langle m|(a+a^\dagger),
\label{eq:Hmulti}
\end{equation}
where
\begin{equation}
\omega_n=\frac{\epsilon_n}{\hbar},
\qquad
g_{nm}=\gamma_B z_{nm}.
\label{eq:gnm}
\end{equation}
Equation~(\ref{eq:Hmulti}) provides the natural starting point for the analysis. The in-plane field sets an overall coupling scale through $\gamma_B$, whereas the actual transition amplitudes are the matrix-element-dressed quantities $g_{nm}$. The model is therefore intrinsically multilevel before any further approximation is introduced.

It is useful to separate the diagonal and off-diagonal parts of the interaction,
\begin{equation}
\begin{split}
\sum_{n,m} g_{nm}\ket{n}\bra{m}(a+a^\dagger)
={}& \sum_n d_n\ket{n}\bra{n}(a+a^\dagger) \\
&+ \sum_{n\neq m} g_{nm}\ket{n}\bra{m}(a+a^\dagger),
\end{split}
\label{eq:diagoff}
\end{equation}

with $d_n=\gamma_B z_{nn}$. The diagonal term produces a subband-dependent displacement of the cyclotron oscillator and contributes only the constant shift $-d_n^2/\omega_c$ after completing the square. The state-dependent cyclotron splitting is generated by the off-diagonal term, which mediates virtual transitions between different vertical subbands.

\paragraph*{Dispersive cyclotron shift.}
We separate the interaction in Eq.~(\ref{eq:diagoff}) into its diagonal and off-diagonal parts and treat the latter to second order. For the product states $|m,\ell\rangle=|m\rangle\otimes|\ell\rangle$ of the uncoupled vertical and cyclotron motions, standard perturbation theory gives the subband-dependent cyclotron shift
\begin{equation}
\begin{aligned}
\delta\omega_c^{(m)}
&=
-\sum_{n\neq m}|g_{nm}|^2
\left(
\frac{1}{\omega_{nm}-\omega_c}
+
\frac{1}{\omega_{nm}+\omega_c}
\right)
\\
&=
-2\sum_{n\neq m}
\frac{|g_{nm}|^2\,\omega_{nm}}
{\omega_{nm}^2-\omega_c^2},
\end{aligned}
\label{eq:dwc}
\end{equation}
where $\omega_{nm}=\omega_n-\omega_m$.
Equation~(\ref{eq:dwc}) includes virtual transitions through the full vertical ladder and forms the basis of the calculations below.

The second-order approximation requires
\begin{equation}
\eta_{nm}^{(\pm)}
=
\frac{|g_{nm}|}{|\omega_{nm}\pm\omega_c|}
\ll 1
\label{eq:etas}
\end{equation}
for the virtual channels that make appreciable contributions to
Eq.~(\ref{eq:dwc}).

\paragraph*{Cyclotron line as a probe of vertical susceptibility.}

Defining the frequency-dependent coordinate susceptibility of the vertical state $|m\rangle$ as

\begin{equation}
\chi_m^{zz}(\omega)=2\sum_{n\neq m}|z_{nm}|^2
\frac{\omega_{nm}}{\omega_{nm}^2-\omega^2},
\label{eq:chi}
\end{equation}
the cyclotron shift becomes
\begin{equation}
\delta\omega_c^{(m)}=-\gamma_B^2\chi_m^{zz}(\omega_c).
\label{eq:dwcpolar}
\end{equation}
The cyclotron resonance therefore probes the dynamical susceptibility of the vertical Rydberg ladder at the frequency $\omega_c$. We refer to this response as the vertical polarizability in the discussion below.

Equation~(\ref{eq:dwc}) describes the shift produced by the tilted-field coupling within the isolated single-electron Hamiltonian. It does not include the additional renormalization of the cyclotron resonance caused by coupling to surface excitations of liquid helium. Such polaronic effects produce a frequency shift and an effective-mass renormalization rather than merely a decay rate, and their magnitude can depend on the pressing field, temperature, and electron density \cite{monarkha2004book,grimes1976cyclotron,edelman1976effective,wilen1988cyclotron, chepelianskii2021magnetoplasmons}. In an experiment, the state-dependent shift calculated here must therefore be resolved relative to a cyclotron resonance whose baseline frequency may already contain polaronic and many-electron corrections.

The cyclotron resonance therefore probes the dynamical polarizability of the vertical Rydberg ladder at the probe frequency $\omega_c$. This is the sense in which the tilted field turns one electron into a self-coupled probe of its own multilevel structure.

For the two lowest vertical states one finds
\begin{align}
\delta\omega_c^{(1)}&=-2\sum_{n>1}\frac{|g_{n1}|^2\omega_{n1}}{\omega_{n1}^2-\omega_c^2},
\label{eq:dwc1}\\
\delta\omega_c^{(2)}&=+2\frac{|g_{21}|^2\omega_{21}}{\omega_{21}^2-\omega_c^2}
-2\sum_{n>2}\frac{|g_{n2}|^2\omega_{n2}}{\omega_{n2}^2-\omega_c^2}.
\label{eq:dwc2}
\end{align}

\begin{figure}[t]
\includegraphics[width=\columnwidth]{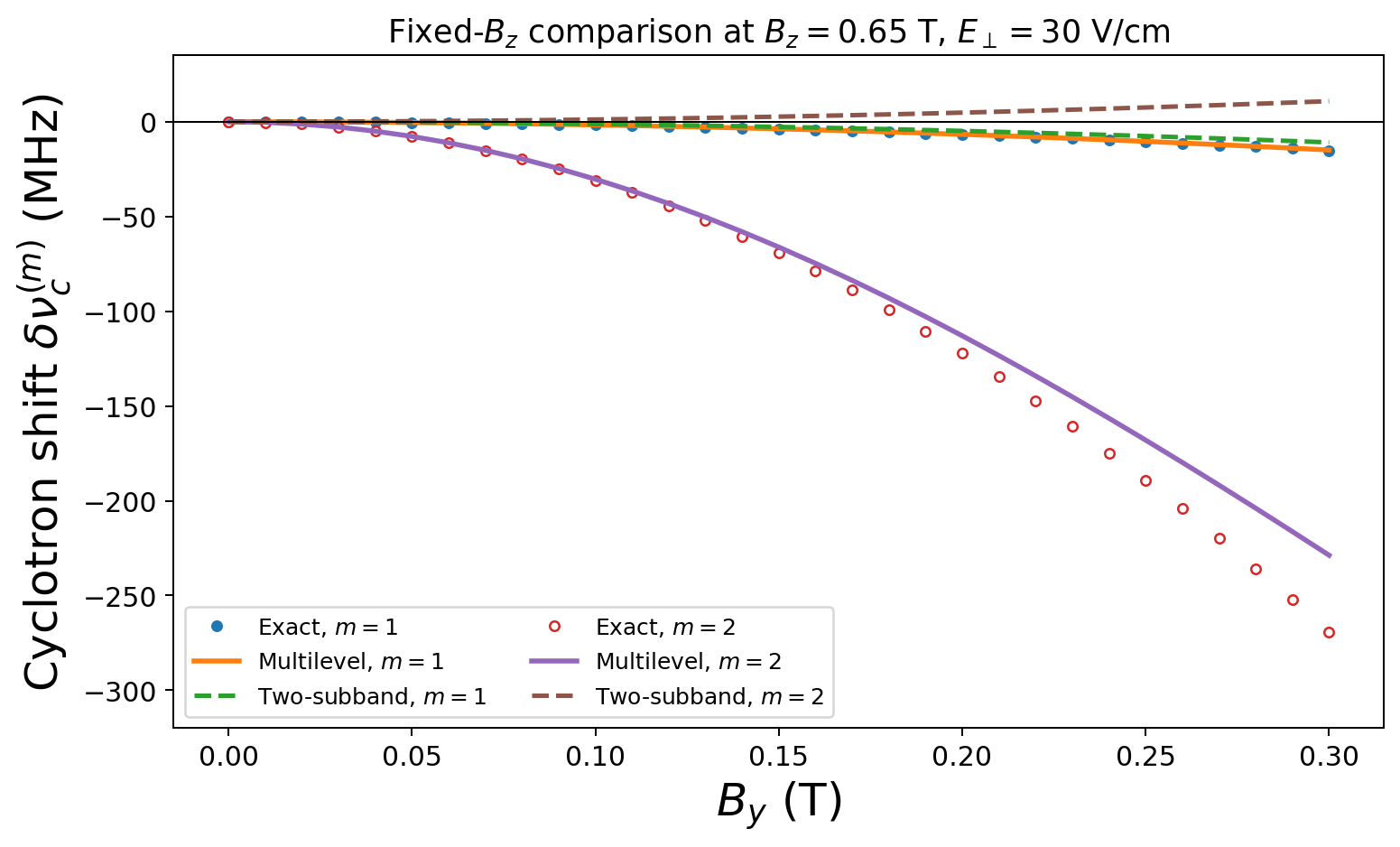}
\caption{
Cyclotron-branch shifts at fixed $B_z=0.65~\mathrm{T}$ and $E_\perp=30~\mathrm{V/cm}$ as a function of the in-plane field $B_y$. For these parameters $\nu_{21}=140.2~\mathrm{GHz}$ at $B_y=0$. Symbols show exact diagonalization of the microscopic Hamiltonian. Solid lines show the multilevel second-order theory of Eqs.~(\ref{eq:dwc1}) and (\ref{eq:dwc2}). Dashed lines show the strict two-subband limit. The ground-state branch is described reasonably well by the two-subband reduction, whereas the first-excited branch departs strongly from it while remaining well captured by the multilevel theory.}
\label{fig:fixed_bz}
\end{figure}

Equations~(\ref{eq:dwc1}) and (\ref{eq:dwc2}) also clarify why the problem is generically multilevel. The ground-state branch is often dominated by the virtual transition $1\leftrightarrow2$, so a two-subband estimate can remain qualitatively reasonable. The first-excited branch is much more sensitive to higher subbands because upward and downward virtual channels enter with opposite signs. The first approximation to lose quantitative accuracy is therefore the two-subband truncation, not the dispersive expansion itself. A channel-resolved decomposition is given in Appendix~\ref{app:convergence}. At the representative point $B_z=0.65~\mathrm{T}$ and $B_y=0.30~\mathrm{T}$, the $n=2$ virtual channel accounts for about $73\%$ of $\delta\omega_c^{(1)}$, whereas the $n=3$ channel accounts for about $95\%$ of $\delta\omega_c^{(2)}$ when the sums are truncated at $N_z=8$. Contributions from higher subbands decrease rapidly on the scale of the shifts shown in Fig.~\ref{fig:fixed_bz}.

\paragraph*{Two-subband limit.}
A projection onto the two lowest vertical subbands yields a generalized Rabi form. With
\begin{equation}
\begin{aligned}
P&=\ket{1}\bra{1}+\ket{2}\bra{2},\\
\tau_z&=\ket{2}\bra{2}-\ket{1}\bra{1},\\
\tau_x&=\ket{1}\bra{2}+\ket{2}\bra{1}.
\end{aligned}
\label{eq:projectors}
\end{equation}
the projected coordinate operator is
\begin{equation}
PzP=z_0 I+\frac{\delta z}{2}\tau_z+z_{12}\tau_x,
\label{eq:PzP}
\end{equation}
where
\begin{equation}
z_0=\frac{z_{11}+z_{22}}{2},
\qquad
\delta z=z_{22}-z_{11}.
\label{eq:z0dz}
\end{equation}
The corresponding two-subband Hamiltonian reads
\begin{equation}
\begin{aligned}
\frac{H_{2\mathrm{sb}}}{\hbar}={}&\frac{\omega_q}{2}\tau_z+\omega_c a^\dagger a+g_0(a+a^\dagger)\\
&+g_z\tau_z(a+a^\dagger)+g_x\tau_x(a+a^\dagger),
\end{aligned}
\label{eq:H2sb}
\end{equation}
with
\begin{equation}
\begin{aligned}
\omega_q=\omega_2-\omega_1,
\qquad
g_0=\gamma_B z_0,
\\
g_z=\gamma_B\frac{\delta z}{2},
\qquad
g_x=\gamma_B z_{12}.
\label{eq:gdefs}
\end{aligned}
\end{equation}
The ordinary transverse Rabi Hamiltonian is thus a reduced limit of the microscopic problem. After removing the displacement generated by $g_0$ and neglecting the longitudinal term proportional to $g_z$, one recovers the familiar transverse form. The full two-subband reduction, however, is a generalized Rabi model with both longitudinal and transverse coupling.

In that reduced transverse limit the cyclotron splitting becomes
\begin{equation}
\begin{aligned}
\Delta\omega_c^{(2\mathrm{sb})}=2\left(\chi_{\mathrm{JC}}+\mu_{\mathrm{BS}}\right),
\\
\chi_{\mathrm{JC}}=\frac{g_x^2}{\omega_q-\omega_c},
\qquad
\mu_{\mathrm{BS}}=\frac{g_x^2}{\omega_q+\omega_c},
\label{eq:twosubshift}
\end{aligned}
\end{equation}
where $\chi_{\mathrm{JC}}$ and $\mu_{\mathrm{BS}}$ denote the Jaynes--Cummings and Bloch--Siegert contributions, respectively. Equation~(\ref{eq:twosubshift}) is a useful two-subband limit, but it is not the general result for Eq.~(\ref{eq:Hmicro}).

Both terms in Eq.~(\ref{eq:twosubshift}) are retained in the numerical estimates below. Since $\omega_q\gg\omega_c$ in the parameter range considered here, the denominators $\omega_q-\omega_c$ and $\omega_q+\omega_c$ are of comparable magnitude.

\paragraph*{Comparison with exact diagonalization and accessible parameter window.}
Unless stated otherwise, the numerical results below are for ${}^4\mathrm{He}$ with $E_\perp=30~\mathrm{V/cm}$. For this choice the lowest vertical transition is $\nu_{21}=140.2~\mathrm{GHz}$ at $B_y=0$, and the diamagnetic term in Eq.~(\ref{eq:Hmicro}) shifts it to $142.8~\mathrm{GHz}$ at $B_y=0.30~\mathrm{T}$. The code used to generate Figs.~1--3 is available in Ref.~\cite{shulga_code_2026}.

Figure~\ref{fig:fixed_bz} compares the multilevel dispersive theory with exact diagonalization of the microscopic Hamiltonian and with the strict two-subband limit at fixed $B_z$. The comparison demonstrates that the multilevel second-order theory remains accurate in the same field range where the strict two-subband reduction already becomes quantitatively inadequate for the branch associated with the first excited vertical state.

The vertical Schr\"odinger problem was discretized on a finite-difference grid of 1600 points over the interval $0<z<0.91~\mu\mathrm{m}$. The helium barrier was implemented as a hard wall at $z=0$, corresponding to the $V_0\rightarrow\infty$ limit for the low-lying states considered here. The main figures retain the lowest $N_z=8$ vertical subbands in the multilevel sums. The exact diagonalization of Eq.~(\ref{eq:Hmulti}) used the same $N_z=8$ vertical states together with $N_c=12$ cyclotron states. At the largest tilt shown in Fig.~\ref{fig:fixed_bz}, namely $B_z=0.65~\mathrm{T}$ and $B_y=0.30~\mathrm{T}$, increasing the diagonalization basis to $N_z=16$ and $N_c=16$, together with grid refinements up a grid refinement to 2400 points over $z_{\max}=1.14~\mu\mathrm{m}$, changes the exact branch shifts by less than $0.4~\mathrm{MHz}$. The multilevel sums themselves change by less than $0.6~\mathrm{MHz}$ when the cutoff is increased from $N_z=8$ to $N_z=16$. The numerical results shown below are therefore converged on the MHz scale relevant for the present spectroscopy estimates; additional convergence and channel-resolved data are given in Appendix~\ref{app:convergence}.

\begin{figure}[t]
\includegraphics[width=\columnwidth]{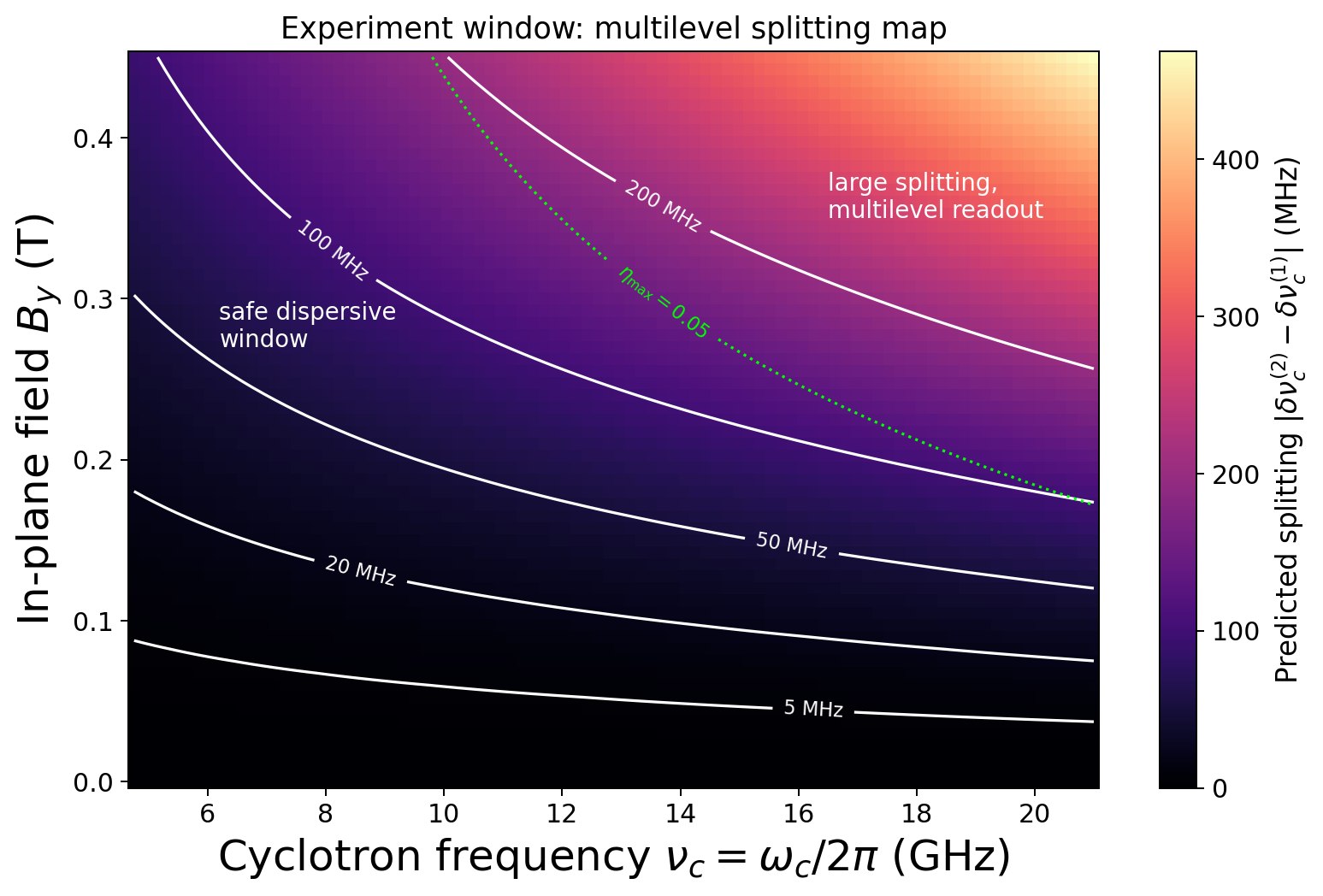}
\caption{
Predicted multilevel cyclotron splitting in the experimentally relevant microwave window. The horizontal axis is the bare cyclotron frequency $\nu_c=\omega_c/2\pi$ and the vertical axis is the in-plane field $B_y$. The color scale gives the splitting $|\delta\nu_c^{(2)}-\delta\nu_c^{(1)}|$ obtained from the multilevel second-order theory for fixed $E_\perp=30~\mathrm{V/cm}$ in ${}^4\mathrm{He}$, corresponding to $\nu_{21}=140.2~\mathrm{GHz}$ at $B_y=0$. The white contours mark 5, 20, 50, 100, and 200~MHz. The dotted contour marks the conservative dispersive boundary $\eta_{\max}=0.05$, based on Eq.~(\ref{eq:etas}).}
\label{fig:window}
\end{figure}

Figure~\ref{fig:window} shows the corresponding splitting map in the experimentally relevant 5-20~GHz cyclotron window. The horizontal axis is the bare cyclotron frequency $\nu_c=\omega_c/2\pi$, set by $B_z$, and the vertical axis is the in-plane field $B_y$, which controls the coupling scale. The color map shows the multilevel splitting $|\delta\nu_c^{(2)}-\delta\nu_c^{(1)}|$, while the contours mark values of direct spectroscopic interest. The figure shows a broad window in which tens to hundreds of MHz of splitting are expected while the detuning criteria in Eq.~(\ref{eq:etas}) still indicate a safely dispersive regime.

Figure~\ref{fig:window} gives the branch separation predicted by the isolated single-electron Hamiltonian. Its experimental visibility depends on the measured cyclotron linewidth and on the preparation of the vertical state. In the weak-probe regime, the two branches can be resolved when
\begin{equation}
\left|
\delta\omega_c^{(2)}-\delta\omega_c^{(1)}
\right|
\gtrsim
\Gamma_{\mathrm{exp}},
\label{eq:resolution}
\end{equation}
where $\Gamma_{\mathrm{exp}}$ denotes the experimentally determined linewidth of the relevant cyclotron or orbital mode. We do not assume a microscopic relaxation model in the present calculation.

\begin{figure*}[t]
\includegraphics[width=\textwidth]{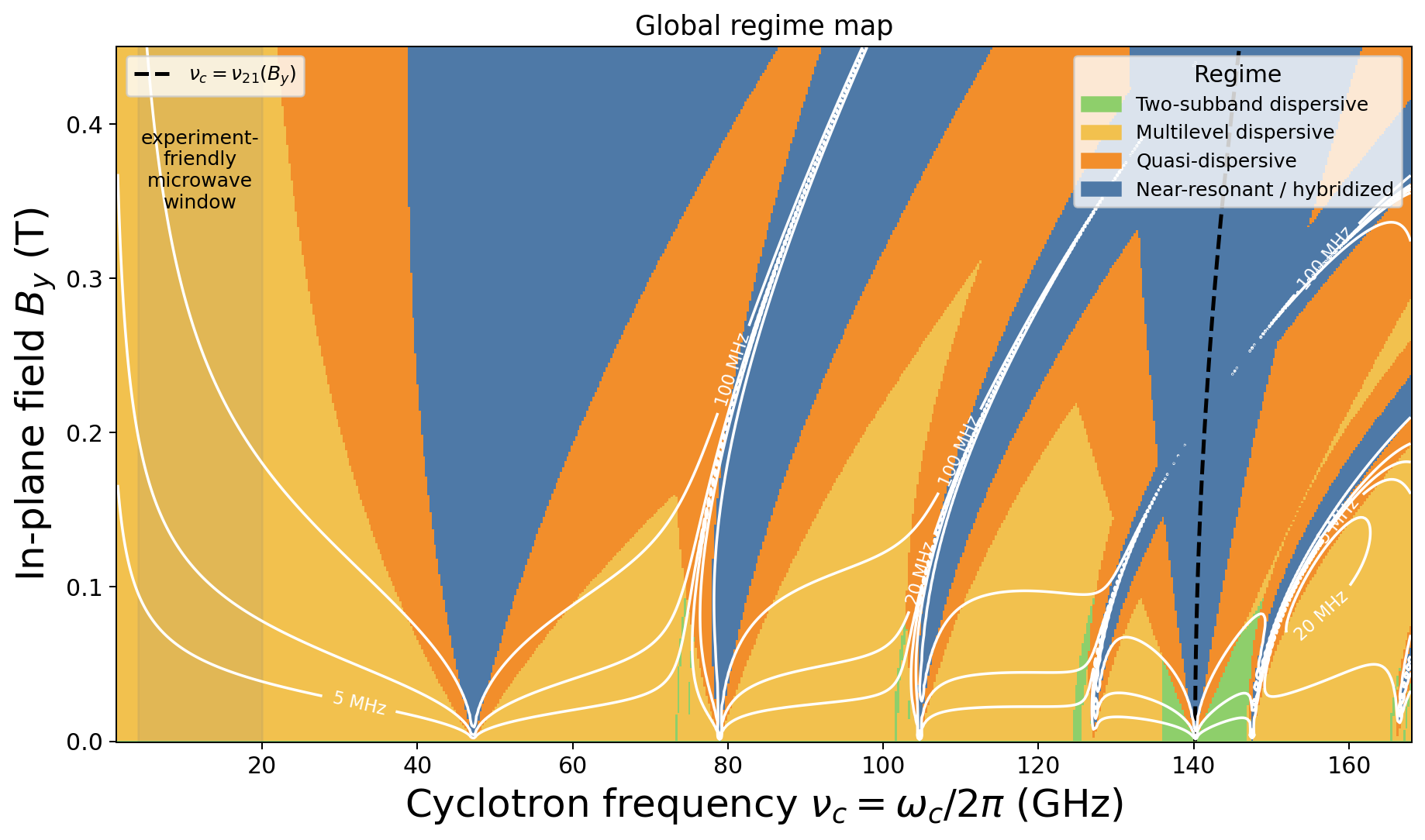}
\caption{
Global regime map of the tilted-field Hamiltonian for ${}^4\mathrm{He}$ at fixed $E_\perp=30~\mathrm{V/cm}$. The horizontal axis is the bare cyclotron frequency $\nu_c=\omega_c/2\pi$ and the vertical axis is the in-plane field $B_y$. The shaded strip indicates the 4-20~GHz microwave window. The dashed black curve marks the condition $\nu_c=\nu_{21}(B_y)$, where the cyclotron frequency matches the lowest vertical transition. White contours show multilevel splittings of 5, 20, and 100~MHz. The green sector corresponds to a strict two-subband dispersive regime. The yellow sector remains dispersive but is already intrinsically multilevel. The orange sector is a quasi-dispersive crossover regime, and the blue sector corresponds to near-resonant or hybridized behavior.}
\label{fig:regimes}
\end{figure*}

The broader structure of the tilted-field Hamiltonian is summarized in Fig.~\ref{fig:regimes}. We classify the parameter space using the maximum detuning ratio $\eta_{\max}$ and the discrepancy between the multilevel and two-subband predictions. The resulting map distinguishes a narrow two-subband dispersive sector from a broader multilevel dispersive sector, followed by a quasi-dispersive crossover and, finally, near-resonant hybridization. The map therefore shows explicitly that the experimentally relevant dispersive window does not coincide with the much narrower region in which a strict two-subband reduction remains quantitatively reliable.

To make the classification used in Fig.~\ref{fig:regimes} explicit, we introduce the quantity
\begin{equation}
\eta_{\max}=
\max_{m\in\{1,2\},\,n\neq m}
\left\{
\frac{|g_{nm}|}{|\omega_{nm}-\omega_c|},
\frac{|g_{nm}|}{|\omega_{nm}+\omega_c|}
\right\},
\label{eq:etamax}
\end{equation}
which measures the proximity of the relevant virtual channels to resonance in the full multilevel problem. In addition, we define the relative deviation of the strict two-subband prediction from the multilevel result,
\begin{equation}
\mathcal{E}_{2\mathrm{sb}}=
\frac{\left|\Delta\omega_c^{(\mathrm{multi})}-\Delta\omega_c^{(2\mathrm{sb})}\right|}{\left|\Delta\omega_c^{(\mathrm{multi})}\right|},
\qquad
\Delta\omega_c^{(\mathrm{multi})}=
\delta\omega_c^{(2)}-\delta\omega_c^{(1)}.
\label{eq:E2sb}
\end{equation}
The parameter $\eta_{\max}$ therefore controls the validity of the dispersive elimination itself, whereas $\mathcal{E}_{2\mathrm{sb}}$ quantifies the additional error introduced by the two-subband truncation. The boundaries shown in Fig.~\ref{fig:regimes} should be understood as operational crossovers rather than sharp phase boundaries: in the map we use $\eta_{\max}<0.10$ as a conservative dispersive criterion, $0.10\le \eta_{\max}<0.25$ as a quasi-dispersive crossover regime, and $\eta_{\max}\ge 0.25$ as a near-resonant or hybridized regime. Within the dispersive sector $\eta_{\max}<0.10$, we further distinguish the strict two-subband regime by requiring $\mathcal{E}_{2\mathrm{sb}}<0.15$; otherwise the point is classified as multilevel dispersive.

The physical content of these regions is different. In the green sector, the dressed eigenstates remain close to product states of a vertical subband and a cyclotron level, and both the dispersive expansion and the two-subband projection are quantitatively accurate. In the yellow sector, the condition $|g_{nm}|\ll |\omega_{nm}\pm\omega_c|$ is still satisfied for the dominant channels, so the cyclotron response remains dispersive, but higher subbands already make an $O(1)$ contribution to the shift and the strict two-subband description is no longer reliable. The orange sector corresponds to a crossover regime in which at least one channel approaches the limit of perturbative elimination; second-order formulas remain qualitatively useful there, but exact diagonalization is required for quantitative work. Finally, in the blue sector one or more denominators $|\omega_{nm}\pm\omega_c|$ become comparable to the corresponding couplings, so the notion of a simple state-dependent cyclotron shift gives way to genuine hybridization of cyclotron and vertical excitations, accompanied by avoided-crossing structure in the spectrum.

\paragraph*{Discussion and experimental outlook.}
The present analysis identifies electrons on helium as a particularly clean realization of a broader class of tilted-field subband--Landau Hamiltonians previously studied in semiconductor structures \cite{ando1979intersubband,merlin1987parabolic,borroff1987raman, observation1988tilted,kumada2008modulation,oh1994threedots}. What distinguishes the helium platform is the combination of strong anharmonicity, realistic single-electron operation, and a far-detuned window in which the cyclotron response remains directly sensitive to the vertical ladder of the same particle. In this regime the microscopic tilted-field Hamiltonian connects the earlier subband--Landau literature with a generalized Rabi description obtained after projection onto the two lowest vertical subbands.

The calculation above assumes translationally invariant in-plane motion and an ideal Landau ladder. For an electrostatically trapped electron, the in-plane eigenfrequencies are combinations of the cyclotron and confinement frequencies, and the coupling must be rederived using the orbital eigenstates of the specific trap. The present results should therefore be regarded as the uniform single-particle limit and as an estimate of the characteristic scale of the state-dependent shift, rather than as a complete model of a particular trapped device.

A direct spectroscopic test of the theory would use a weak pulse on the vertical transition to prepare a controllable population of the first excited subband, followed by a probe of the cyclotron absorption. The predicted signal is not only a split cyclotron line, but a branch evolution that follows the multilevel theory rather than the strict two-subband limit, particularly as $B_y$ is increased at fixed $B_z$. Figure~\ref{fig:fixed_bz} indicates a parameter range in which this distinction should be experimentally resolvable.

A complementary experimental handle is the holding field $E_\perp$. Varying $E_\perp$ at fixed $\nu_c$ in the 4--20~GHz window changes both the vertical detunings $\omega_{nm}$ and the matrix elements $z_{nm}$ entering Eq.~(\ref{eq:dwc}). A measurement of the cyclotron shift as a function of $E_\perp$ would therefore provide a direct test of the susceptibility interpretation in Eq.~(\ref{eq:dwcpolar}) and would distinguish the full multilevel response from a simple $g_{12}^2/\Delta$ scaling law.

A further spectroscopy experiment would follow the evolution from the far-detuned regime to genuine hybridization by sweeping $B_z$ and $B_y$ toward the condition $\nu_c=\nu_{21}(B_y)$. Such a measurement would connect the present far-detuned theory directly to the near-resonant tilted-field studies of Refs.~\cite{yunusova2019coupling,zadorozhko2021motional} and would provide an experimental benchmark for the operational crossover boundaries introduced in Fig.~\ref{fig:regimes}.

The same effective Hamiltonian also motivates a time-domain experiment, but the simple conditional coherent-state picture applies only in a reduced limit. Specifically, the dynamics must first be projected onto the two lowest vertical subbands, the trivial displacement generated by $g_0$ in Eq.~(\ref{eq:H2sb}) must be removed, and the longitudinal term proportional to $g_z$ must remain negligible over the timescale of the measurement. In that reduced two-subband dispersive limit, an initial state
\begin{equation}
\frac{|1\rangle+|2\rangle}{\sqrt{2}}\otimes|\alpha\rangle
\label{eq:initstate_reduced}
\end{equation}
evolves into
\begin{equation}
\begin{aligned}
\frac{|1\rangle|\alpha_1(t)\rangle+e^{-i\phi(t)}|2\rangle|\alpha_2(t)\rangle}{\sqrt{2}},
\\
\alpha_m(t)=\alpha e^{-i[\omega_c+\delta\omega_c^{(m)}]t}.
\label{eq:entstate_reduced}
\end{aligned}
\end{equation}
The overlap of the conditional coherent states,
\begin{equation}
\begin{aligned}
\left|\langle\alpha_2(t)|\alpha_1(t)\rangle\right|
=
\exp\!\left[-2|\alpha|^2\sin^2\!\left(\frac{\Delta\omega_c t}{2}\right)\right],
\\
\Delta\omega_c=\delta\omega_c^{(2)}-\delta\omega_c^{(1)},
\label{eq:overlap_reduced}
\end{aligned}
\end{equation}
directly governs the Ramsey visibility of the vertical two-state subsystem. Outside this reduced limit, the multilevel dispersive regime still leads to state-dependent Ramsey and Rabi signals, but the evolution is no longer described by a single pair of conditional coherent states. In contrast to the familiar circuit-QED setting, the entangling degree of freedom is here not an external cavity mode, but the cyclotron motion of the same electron.

%The same effective Hamiltonian also motivates a time-domain experiment. In the two-subband dispersive limit, an initial state
%\begin{equation}
%\frac{|1\rangle+|2\rangle}{\sqrt{2}}\otimes|\alpha\rangle
%\label{eq:initstate}
%\end{equation}
%evolves into
%\begin{equation}
%\frac{|1\rangle|\alpha_1(t)\rangle+e^{-i\phi(t)}|2\rangle|\alpha_2(t)\rangle}{\sqrt{2}},
%\qquad
%\alpha_m(t)=\alpha e^{-i[\omega_c+\delta\omega_c^{(m)}]t}.
%\label{eq:entstate}
%\end{equation}
%The cyclotron mode then acts as an internal which-path detector for the vertical state of the same electron. The overlap of the conditional coherent states,
%\begin{equation}
%\left|\langle\alpha_2(t)|\alpha_1(t)\rangle\right|
%=\exp\!
%\left[-2|\alpha|^2\sin^2\!\left(\frac{\Delta\omega_c t}{2}\right)\right],
%\label{eq:overlap}
%\end{equation}
%with $\Delta\omega_c=\delta\omega_c^{(2)}-\delta\omega_c^{(1)}$, directly controls the Ramsey visibility of the vertical two-state subsystem. 

%The corresponding reduced purity of the vertical subsystem is
%\begin{equation}
%\mathcal{P}_q(t)=\mathrm{Tr}\,\rho_q^2(t)=\frac12\left[1+\left|\langle\alpha_2(t)|\alpha_1(t)\rangle\right|^2\right].
%\label{eq:purity}
%\end{equation}
A time-domain experiment of this type would therefore probe not only a static splitting, but also the entanglement generated between two internal degrees of freedom of the same electron.

For an extended electron layer, increasing the electron number enhances the absorption signal, but the measured response is no longer that of an isolated electron. Electron--electron interactions, ripplon dressing, and density-dependent broadening modify both the position and the width of the cyclotron resonance \cite{monarkha2004book,edelman1976effective,wilen1988cyclotron, chepelianskii2021magnetoplasmons}. The results of Figs.~\ref{fig:fixed_bz}--\ref{fig:regimes} should consequently be interpreted as single-particle branch shifts. In an ensemble measurement, these shifts would have to be identified on top of the experimentally calibrated polaronic and collective cyclotron response. A quantitative density range cannot be specified independently of the sample geometry and the collective mode being probed.

\paragraph*{Conclusion.}
We have shown that a tilted magnetic field realizes, for a surface-state electron on helium, a self-coupled multilevel Landau-subband Hamiltonian. In the far-detuned regime, the cyclotron line probes the dynamical polarizability of the vertical Rydberg ladder of the same electron.
The second-order cyclotron shift is intrinsically multilevel because several low-lying virtual transitions can contribute appreciably. A generalized Rabi description emerges only after projection onto the two lowest subbands. Exact diagonalization confirms a broad microwave window in which sizable state-dependent cyclotron shifts are expected while the system remains dispersive. These results provide a concrete spectroscopic target and a starting point for future studies of geometry-specific trapped-electron systems and many-body tilted-field response.

\section{Acknowledgements}
This work was supported by the KAKENHI grant (EC-23/22K13985). 

\section*{Data availability}
The numerical code and data used to generate the main figures are available in Ref.~\cite{shulga_code_2026}. The additional notebook used for the convergence and channel-resolved calculations in Appendix~\ref{app:convergence} is available in Ref.~\cite{shulga_revision_code_2026}.

\clearpage
\appendix

\section{Convergence and channel-resolved contributions}
\label{app:convergence}

\begin{figure}[t]
\includegraphics[width=\columnwidth]{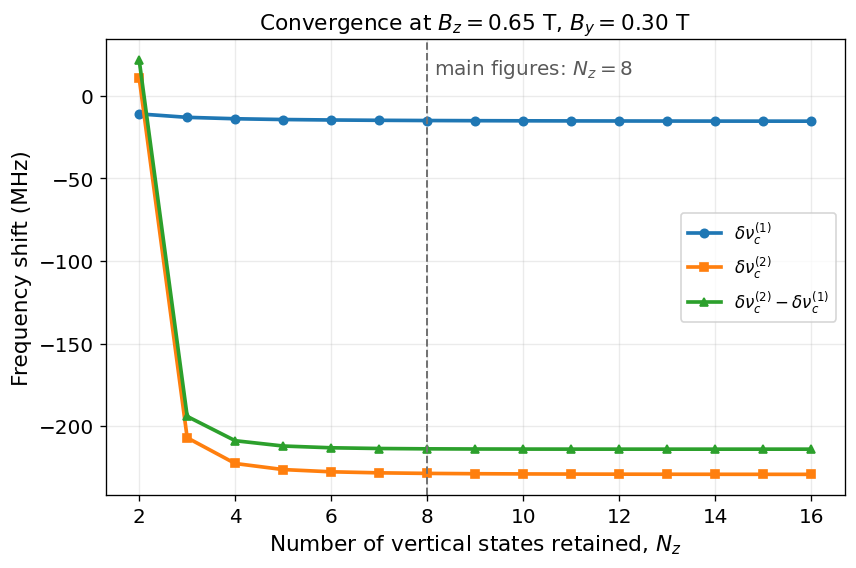}
\caption{Convergence of the multilevel dispersive shifts with the number of vertical states retained in Eq.~(\ref{eq:dwc}), evaluated at
$B_z=0.65~\mathrm{T}$, $B_y=0.30~\mathrm{T}$, and $E_\perp=30~\mathrm{V/cm}$.
The vertical dashed line marks $N_z=8$, the cutoff used for the main figures.
The rightmost points show the values obtained at $N_z=16$.
Increasing the cutoff from $N_z=8$ to $N_z=16$ changes the individual branch shifts by less than $0.6~\mathrm{MHz}$ and the differential splitting by less than $0.2~\mathrm{MHz}$.}
\label{fig:conv_Nz}
\end{figure}

\begin{figure}[t]
\includegraphics[width=\columnwidth]{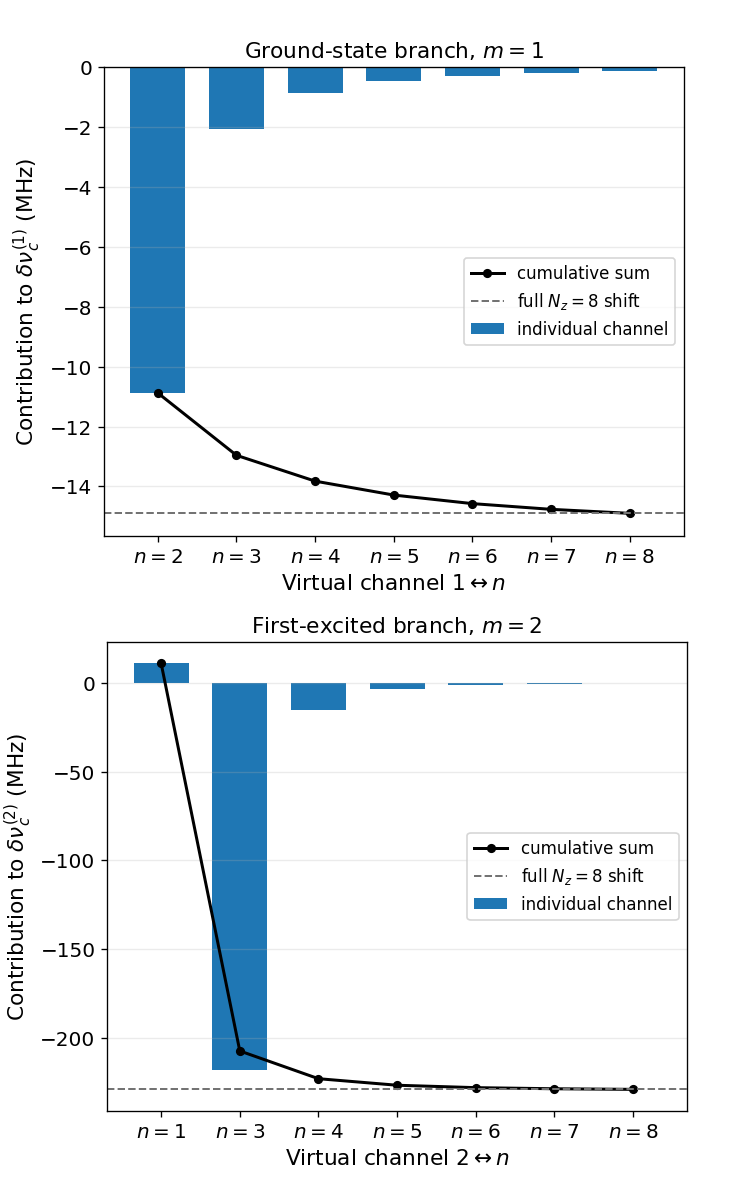}
\caption{Channel-resolved decomposition of the multilevel cyclotron shifts at $B_z=0.65~\mathrm{T}$, $B_y=0.30~\mathrm{T}$, and $E_\perp=30~\mathrm{V/cm}$. Bars show individual virtual-channel contributions to Eqs.~(\ref{eq:dwc1}) and (\ref{eq:dwc2}); black markers show cumulative partial sums. Top: ground-state branch. Bottom: first-excited branch.}
\label{fig:channels}
\end{figure}

This appendix records the numerical convergence checks used to support the results in Fig.~\ref{fig:fixed_bz}. The checks are performed at the largest tilt shown in that figure, $B_z=0.65~\mathrm{T}$ and $B_y=0.30~\mathrm{T}$, where convergence is the most demanding in the plotted range. The notebook used to generate Figs.~\ref{fig:conv_Nz} and \ref{fig:channels} is available in Ref.~\cite{shulga_revision_code_2026}.

Figure~\ref{fig:conv_Nz} shows the multilevel shifts as a function of the number of vertical states retained in Eq.~(\ref{eq:dwc}). Increasing the cutoff from $N_z=8$ to $N_z=16$ changes $\delta\nu_c^{(1)}$ by $0.39~\mathrm{MHz}$ and $\delta\nu_c^{(2)}$ by $0.58~\mathrm{MHz}$. The corresponding differential splitting changes by $0.19~\mathrm{MHz}$. This is small compared with the tens-to-hundreds of MHz shifts discussed in the main text.

Figure~\ref{fig:channels} gives the corresponding channel-resolved decomposition. For the ground-state branch, the $1\leftrightarrow2$ channel gives the dominant contribution. For the first-excited branch, the largest term is the upward $2\leftrightarrow3$ virtual channel, whereas the $2\leftrightarrow1$ term enters with the opposite sign and partially compensates it. The higher-subband contribution is therefore important, but it is not a slowly convergent tail of many high-lying states; the deviation from the two-subband model is dominated by a small number of low-lying virtual channels.

\bibliographystyle{apsrev4-2}
\bibliography{main_revised_response}

\begin{center}
\small
\textcopyright{} 2026 American Physical Society. 
This is the accepted manuscript of the following article:

\medskip

K.~Shulga, ``Internal cyclotron probe of the vertical polarizability 
of a surface-state electron on helium,'' 
\textit{Phys. Rev. A} \textbf{114}, 023708 (2026).

\medskip

The final published version is available at 
\href{https://doi.org/10.1103/96p7-vj9f}
{https://doi.org/10.1103/96p7-vj9f}. 
This manuscript version is posted with permission for 
non-commercial scholarly use.
\end{center}

\end{document}